\documentclass[times,10pt,onecolumn]{article}
\usepackage[top=4.3pc, bottom=4.3pc, left=4.5pc, right=4.5pc]{geometry}
\usepackage[linesnumbered,ruled,vlined,titlenotnumbered,noresetcount]{algorithm2e}
\usepackage{csquotes,url,multirow,multicol,subcaption,array,amssymb,mathtools,wasysym,stmaryrd,pifont}

\usepackage{balance,lipsum,t1enc,anyfontsize,amsmath,amsfonts,amsthm,graphicx,color,wrapfig}
\usepackage[warn]{textcomp}
\usepackage[shortlabels]{enumitem}
\usepackage[table]{xcolor}
\usepackage[title,titletoc,toc]{appendix}
\usepackage{cleveref,soul}
\usepackage{authblk,titlesec}
\crefname{algocf}{alg.}{algs.}
\Crefname{algocf}{Algorithm}{Algorithms}
\crefname{AlgoLine}{line}{lines}
\Crefname{AlgoLine}{Line}{Lines}
\SetAlgoCaptionSeparator{.}

\theoremstyle{plain}

\newtheorem{proposition}{{\sc Proposition}}

\newtheorem{definition}{{\sc Definition}}
\newtheorem{lemma}{{\sc Lemma}}
\newtheorem{cond}{{\sc Condition}}
\theoremstyle{definition}

\fontsize{5}{6}\selectfont
\let\oldnl\nl
\newcommand{\nonl}{\renewcommand{\nl}{\let\nl\oldnl}}

\titlespacing*{\section}
{0pt}{2ex plus 1ex minus .2ex}{1ex plus .2ex}
\titlespacing*{\subsection}
{0pt}{1.9ex plus 1ex minus .2ex}{0.9ex plus .2ex}
\titlespacing*{\subsubsection}
{0pt}{1.5ex plus 1ex minus .2ex}{0.7ex plus .2ex}

\begin{document}

\title{Efficient Lock-free Binary Search Trees\footnote{Published as Technical Report at Chalmers University of Technology TR - 2014:05, ISSN 1652-926X in February 2014 and submitted to PODC 2014}}

\author{Bapi Chatterjee, Nhan Nguyen and Philippas Tsigas\\
  Department of Computer Science and Engineering\\
  Chalmers University of Technology, \\
  Gothenburg, Sweden \\
  \texttt{\{bapic,nhann,tsigas\}@chalmers.se}
}
\date{}
\maketitle

\SetNoFillComment
\SetAlgoSkip{}
\SetKw{bl}{bool}
\SetKw{vd}{void}
\SetKw{kt}{KeyType}
\SetKw{el}{KType}
\SetKw{integer}{int}
\SetKw{dat}{Data}
\SetKw{nod}{Node}
\SetKw{lptr}{LocPtr}
\SetKw{dptr}{DataPtr}
\SetKw{nptr}{NPtr}
\SetKw{nd}{Node}
\SetKw{ret}{return}
\SetKw{cont}{continue}
\SetKw{brk}{break}
\SetKw{new}{new}
\SetKw{gt}{goto}
\SetKw{an}{and}
\SetKw{orr}{or}
\SetKw{tdfstr}{struct}
\SetKwFunction{cmp}{cmp}
\SetKwFunction{itrd}{isThreaded}
\SetKwFunction{imkd}{isMarked}
\SetKwFunction{mkd}{getMarked}
\SetKwFunction{umkd}{unMarked}
\SetKwFunction{fgd}{getFlagged}
\SetKwFunction{ifgd}{isFlagged}
\SetKwData{loc}{Location}
\SetKw{data}{data}
\SetKwData{ky}{key}
\SetKwData{e}{k}
\SetKwData{bacnod}{backNode}
\SetKwData{pdir}{pDir}
\SetKwData{dir}{dir}
\SetKwData{mdir}{markDir}
\SetKwData{cur}{curr}
\SetKwData{pre}{prev}
\SetKwData{back}{back}
\SetKwData{isth}{isThread}
\SetKwData{lc}{locAddr}
\SetKwData{fd}{finishDir}
\SetKwData{mv}{mover}
\SetKwData{tru}{true}
\SetKwData{fal}{false}
\SetKwData{nul}{null}
\newcommand{\spsym}{\hspace{-0.3em}{\shortrightarrow}}
\newcommand{\psym}{{\shortrightarrow}}
\newcommand{\smsym}{\hspace{-0.3em}{\cdot}}



\begin{abstract}
In this paper we present a novel algorithm for concurrent lock-free internal binary search trees (BST) and implement a Set abstract data type (ADT) based on that. We show that in the presented lock-free BST algorithm the  amortized step complexity of each set operation - {\sc Add}, {\sc Remove} and {\sc Contains} - is $O(H(n) + c)$, where, $H(n)$ is the height of BST with $n$ number of nodes and $c$ is the contention during the execution. Our algorithm adapts to contention measures according to read-write load. If the situation is read-heavy, the operations avoid helping pending concurrent {\sc Remove} operations during traversal, and, adapt to interval contention. However, for write-heavy situations we let an operation help pending {\sc Remove}, even though it is not obstructed, and so adapt to tighter point contention. It uses single-word compare-and-swap (\texttt{CAS}) operations. We show that our algorithm has improved disjoint-access-parallelism compared to similar existing algorithms. We prove that the presented algorithm is linearizable.  To the best of our knowledge this is the first algorithm for any concurrent tree data structure in which the modify operations are performed with an additive term of contention measure.
\end{abstract}
{\scriptsize {\bf Keywords}: concurrent data structures, binary search tree, amortized analysis, shared memory, lock-free, CAS\\}

\section{Introduction}\label{secIntro}
\indent With the wide and ever-growing availability of multi-core processors it is evermore compelling to design and develop more efficient concurrent data structures. The non-blocking concurrent data structures are more attractive than their blocking counterparts because of obvious reasons of being immune to deadlocks due to various fault-causing factors beyond the control of the data structure designer. Lock-free data structures provide guarantee of deadlock/livelock freedom with fault tolerance and are usually faster than wait-free ones which provide additional guarantee of starvation freedom at the cost of increased programming complexity.

\indent In literature, there are lock-free as well as wait-free singly linked-lists~\cite{fomitchev2004lock,Timmet2012}, lock-free doubly linked-list~\cite{sundell2005lock}, lock-free hash-tables~\cite{Michael200273} and lock-free skip-lists~\cite{fomitchev2004lock,Sundell2005609}. However, not many performance-efficient non-blocking concurrent search trees are available. A multi-word compare-and-swap (\texttt{MCAS}) based lock-free BST implementation was presented by Fraser in~\cite{fraser2004practical}. However, \texttt{MCAS} is not a native atomic primitive provided by available multi-core chips and is very costly to be implemented using single-word \texttt{CAS}. Bronson et al. proposed an optimistic lock-based partially-external BST with relaxed balance~\cite{bronson2010practical}. Ellen et al. presented lock-free external binary search tree~\cite{Ellen:2010:NBS:1835698.1835736} based on co-operative helping technique presented by Barnes~\cite{barnes1993method}. Though their work did not include an analysis of complexity or any empirical evaluation of the algorithm, the contention window of update operations in the data structure is large. Also, because it is an external binary search tree, {\sc Delete} are simpler at the cost of extra memory to maintain internal nodes without actual values. Howley et al. presented a non-blocking internal binary search tree~\cite{Howley:2012:NIB:2312005.2312036} based on similar technique. A software transactional memory based approach was presented by Crain et al.~\cite{DBLP:conf/ppopp/CrainGR12} to design a concurrent red-black tree. While this approach seems to outperform some coarse-grained locking methods, they are easily vanquished by a carefully tailored locking scheme as in \cite{bronson2010practical}. Recently, two lock-free external BSTs~\cite{Natarajan_2014_FCL,Brown_2014_GTN} and a lock-based internal BST~\cite{Drachsler_2014_PCB} have been proposed. All of these works lack comprehensive theoretical complexity analysis.

\indent In an internal BST, {\sc Add} operations start from the root and finish at a leaf node, where the new element is inserted. To {\sc Remove} a node which has both the children present, its successor or predecessor is shifted to take its place. A common predicament for the existing lock-free BST algorithms is that if multiple modify operations contend at a leaf node, and, if a {\sc Remove} operation among them succeeds then all other operations have to restart from the root. It results in the complexity of a modify operation to be $O(cH(n))$ where $H(n)$ is the height of the BST on $n$ nodes and $c$ is the measure of contention. It may grow dramatically with the growth in the size of the tree and the contention. In addition to that, {\sc Contains} operations have to be aware of ongoing {\sc Remove} of nodes with both children present, otherwise, it may return invalid results, and, hence in the existing implementations of lock-free internal BST~\cite{Howley:2012:NIB:2312005.2312036}, they also may have to restart from the root on realizing that the return may be invalid if the other subtree is not scanned. The external or partially-external BSTs remain immune to the latter problem at a cost of extra memory for the routing internal nodes.  Our algorithm solves both these problems elegantly. The {\sc Contains} operations in our BST enjoy oblivion of any kind of modify operation as long as we do not put them to help a concurrent {\sc Remove}, which may be needed only in write-heavy situations. Also, the modify operations after helping a concurrent modify operation restart not from the root rather from a level in the vicinity of failure. It ensures that all the operations in our algorithm run in $O(H(n) + c)$. This is our main contribution.

\indent We always strive to exploit maximum possible disjoint-access-parallelism~\cite{israeli1994disjoint} in a concurrent data structure. The lock-free methods for BST \cite{Ellen:2010:NBS:1835698.1835736, Howley:2012:NIB:2312005.2312036}, in order to use single-word \texttt{CAS} for modifying the outgoing links atomically, and yet maintain correctness, store a flag as an operation field or some version indicator in the node itself, and hence a modify operation \enquote{holds} a node. This way of holding a node, specifically for a \textsc{Remove}, can reduce the progress of two operations which may remain non-conflicting if modifying two outgoing links concurrently. In~\cite{Natarajan_2014_FCL}, a flag is stored in a link instead of a node in an external BST. We chase this problem of just holding a link for an internal BST. We find that it is indeed possible that a \textsc{Remove} operation, instead of holding the node, just holds the links connected to and from a node in a determined order so that maximum possible progress of two concurrent operations, working at two disjoint memory words corresponding to two links can be ensured. We take the business of ``storing a flag'' to the link level from the node level which significantly improves the disjoint-access-parallelism. This is our next contribution.

\indent Helping mechanism which ensures non-blocking progress may prove counterproductive to the performance if not used judiciously. However, at times, if the proportion of {\sc Remove} operations increases, which may need help to finish their pending steps, it is better to help them, so that the traversal path does not contain large number of ``under removal'' nodes. Keeping that in view, we take helping to a level of adaptability to the read-write load: we provide choice over whether an operation, during its traversal, helps an ongoing {\sc Remove} operation. We believe that this adaptive conservative helping in internal BSTs may be very useful in some situations. This is a useful contribution of this work.

\indent Our algorithm requires only single-word atomic \texttt{CAS} primitives along with single-word atomic read and write which now exist in practically all the widely available multi-core processors in the market. Based on our design, we implement a Set ADT. We prove that our algorithm is linearizable~\cite{herlihy1990linearizability}. We also present complexity analysis of our implementation which is lacking in existing lock-free BST algorithms. This is another contribution in this paper.

\indent The body of our paper will further consist of the following sections. In section \ref{secExistStruct}, we present the basic tree terminologies. In section \ref{secAlgo}, the proposed algorithm is described. Section \ref{secAnalysis} presents a discussion on the correctness and the progress of our concurrent implementation along with an amortized analysis of its step complexity. The paper is concluded in section \ref{secConclude}.

\section{Preliminaries}\label{secExistStruct}
A \textit{binary tree} is an ordered tree in which each \textit{node} $x$ has a \textit{left child} and a \textit{right child} denoted as $left(x)$ and $right(x)$ respectively, either or both of which may be \textit{external}. When both the children are external the node is called a \textit{leaf}, with one external child a \textit{unary} node and with no external child a \textit{binary} node, and, all these non-external nodes are called \textit{internal} nodes.
 We denote the parent of a node $x$ by $p(x)$ and there is a unique node called \textit{root} s.t. $p(root) = null$. Each parent is connected with its children via pointers as links (we shall be often using the term pointer and link interchangeably when the context will be understood).
 We are primarily interested in implementing an ordered \textit{Set} ADT - \textit{binary search tree} using a binary tree in which each node is associated with a unique key $k$ selected from a totally ordered universe. A node with a key $k$ is denoted as $x(k)$ and $x$ if the context is otherwise understood. Determined by the total order of the keys, each node $x$ has a \textit{predecessor} and a \textit{successor}, denoted as $pre(x)$ and $suc(x)$, respectively. We denote height of $x$ by $ht(x)$, which is defined as the distance of the deepest leaf in the subtree rooted at $x$ from $x$.\\
\indent We focus on \textit{internal BSTs}, in which, all the internal nodes are \textit{data-nodes} and the external nodes are usually denoted by \nul. There is a \textit{symmetric order} of arranging the data - all the nodes in the \textit{left subtree} of $x(k)$ have keys less than $k$ and all the nodes in its \textit{right subtree} have keys greater than $k$, and so no two nodes can have the same key. To query if the BST {\sc Contains} a data with key $k$, at every \textit{search-step} we utilize this order to look for the desired node either in the left or in the right subtree of the \textit{current node} if not matched at it, unless we reach an external node. On reaching an external node we return \fal, else, if the key matches at a node then we return \tru or address of the node if needed.
To {\sc Add} data, we query by its key $k$. On reaching an external node we replace this node with a new leaf node $x(k)$.
To {\sc Remove} a data-node corresponding to key $k$ we check whether $x(k)$ is in the BST. If the BST does not contain $x(k)$, \fal is returned. On finding a node with key $k$, we perform delete as following. If it is a leaf then we just replace it with an external node. In case of a unary node its only child is connected to its parent.
 For a binary node $x$, it is first replaced with $pre(x)$ or $suc(x)$, which may happen to be a leaf or a unary node, and then the replacer is removed.\\
\indent In an alternate form - an \textit{external BST}, all the internal nodes are \textit{routing-nodes} and the external nodes are data-nodes. In this paper we focus on internal BSTs, and hence forward, by a BST we shall mean an internal BST. 
\section{Our Algorithm}\label{secAlgo}
\subsection{The Efficient Lock-free BST Design}

\begin{figure}[t!]
\centering
    \includegraphics[width=0.48\textwidth]{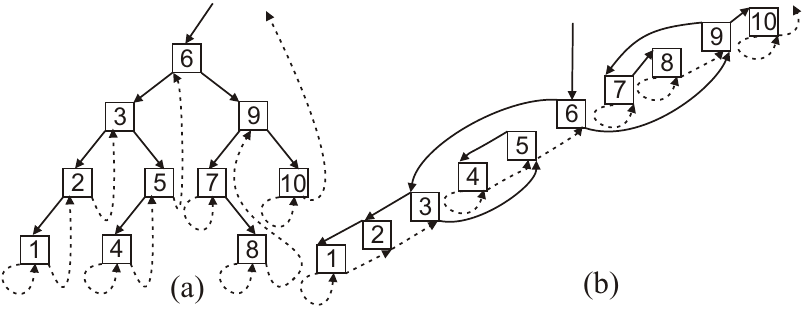}
        \caption{(a) Threaded BST (b) Equivalent ordered list.}\label{bst}
\end{figure}

To implement a lock-free BST, we represent it in a threaded format \cite{perlis1960symbol}. In this format, if the left or the right child pointers at $x$ is null and so corresponds to an external node, it is instead connected to $pre(x)$ or $suc(x)$, respectively. Some indicator is stored to indicate whether a child-link is used for such a connection. This is called \textit{threading} of the child-links. In our design, we use the null child pointers at the leaf and unary nodes as following - right child pointer, if null, is threaded and is used to point to the successor node, whereas, a similar left child pointer is threaded to point to the node itself, see Fig. \ref{bst}(a). In this representation a binary tree can be viewed as an ordered list with exactly two outgoing and two incoming pointers per node, as shown in Fig. \ref{bst}(b). Also, among two incoming pointers, exactly one is threaded and the other is not. Further, if $x(k_i)$ and $x(k_j)$ are two nodes in the BST and there is no node $x(k)$ such that $k_i{\leq}k{\leq}k_j$ then the interval $[k_i,~k_j]$ is called \textit{associated} with the threaded link incoming at $k_j$.

\indent We exploit this symmetry of the equal number of incoming and outgoing pointers. A usual traversal in the BST following its symmetric order for a predecessor query, is equivalent to a traversal over a \textit{subsequence} of the ordered-list produced by an in-order traversal of the BST, which is exactly the one shown in Fig. \ref{bst}(b). This is made possible by the threaded right-links at leaf or unary nodes. Though in this representation, there are two pointers in both incoming and outgoing directions at each node, a single pointer needs to be modified to {\sc Add} a node in the list. To {\sc Remove} a node we may have to modify up to four pointers. Therefore, {\sc Add} can be as simple as that in a lock-free single linked-list~\cite{fomitchev2004lock}, and {\sc Remove} is no more complex than that in a lock-free double linked-list~\cite{sundell2005lock}.  A traversal in a lock-free list may enjoy oblivion from a concurrent {\sc Remove} of a node. Also in our design of internal BST, a traversal can remain undeterred by any ongoing modification, unlike that in existing lock-free implementations of internal BSTs~\cite{Howley:2012:NIB:2312005.2312036,fraser2004practical}.

\begin{figure}[t!]
\centering
        \includegraphics[width=0.48\textwidth]{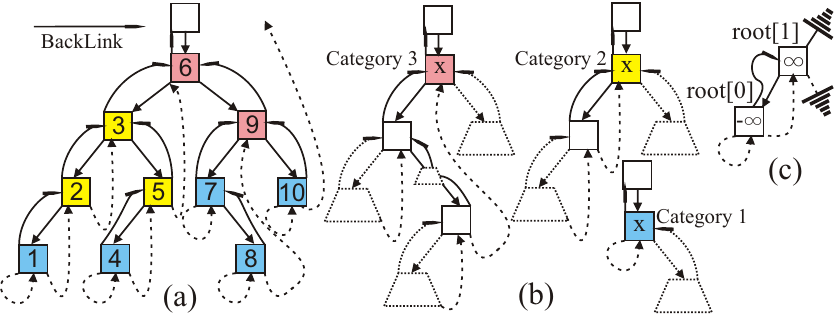}
        \caption{(a) Threaded BST with backlinks (b) Categorization of nodes for {\sc Remove}  (c) An empty tree.}\label{bst_node_category}
\end{figure}

\indent In all the existing designs of lock-free BSTs, when an operation fails at a link connecting to an external node because of a concurrent modify operation, it retries from the scratch i.e. it restarts the operation from the root of the tree, after helping the obstructing concurrent operation. We aim to avoid the ``retry from scratch'' behavior and rather restart from a node at the vicinity of the link where the operation fails, after the required helping. To achieve that, we need to get hold of the appropriate node(s) to restart the failed operations. We use a \textit{backlink} per node which is guaranteed to point to a node \textit{present} in the tree from where the failure spot is always a single link away. It should be noted that a backlink is not used for a tree traversal, see Fig. \ref{bst_node_category}(a).

\indent In designing an internal BST in a concurrent setup, the most difficult part is to perform an error-free {\sc Remove} of a binary node. To remove a binary node we replace it with its predecessor, and hence, the incoming and outgoing links of the predecessor also need to be modified in addition to the incoming and outgoing links of the node itself. According to the number of links needed to be modified in order to remove a node, unlike traditional categorization of nodes of a BST into leaf, unary and binary, we categorize them into three categories as shown in Fig. \ref{bst_node_category}(b). The categorization characteristic is the origin of the threaded incoming link into the node, hereafter we call this link the \textit{order-link}. Nodes belonging to category 1 are those whose order-link emanates from themselves; to category 2, it emanates from the left-child of the node, and to category 3 are those whose incoming order-link emanates from a ``distant'' node in its left-subtree. We name the node where the order-link emanates from, an \textit{order-node}.

\indent To remove a node of category 1, only the incoming parent-link needs to be modified to connect to the node pointed by the right-link. For a category 2, the parent-link is updated to connect to the node pointed by the left-link and the order-link is modified to point to the node which the right-link was pointing to. In order to remove a category 3 node, its predecessor replaces it and the incoming and outgoing links of the predecessor are updated to take the values of that of the removed node. Parent-link of the predecessor is connected to the node which its left-link was pointing to before it got shifted. Also, when a link is updated the thread indicator value of the link, which it updates to, is copied to it. Please note that in this categorization, a unary or a binary node whose left child is left-unary (i.e. whose right child is \nul) or a leaf, gets classified in to category 2. Whereas, category 1 includes conventional leaf and right-unary nodes (whose left child is \nul).

\begin{figure}[t!]
\centering
        \includegraphics[width=0.48\textwidth]{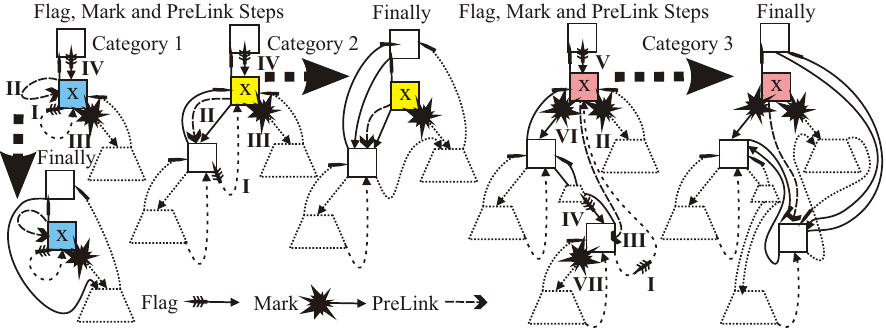}
        \caption{Remove steps of nodes}\label{bst_del}
\end{figure}

\indent Having categorized the nodes as above we describe the modification of links associated with a node undergoing {\sc Remove}. Storing an indicator bit in a pointer has been used in many previous papers~\cite{harris2001pragmatic, fomitchev2004lock, sundell2005lock, Natarajan_2014_FCL}. We use a similar technique as in~\cite{fomitchev2004lock}. Before swapping the pointers associated with a node we flag the incoming pointers to the node and its possible predecessor and mark the outgoing pointers from the same. Note that because a backlink is neither used for traversal nor for injecting a modify operation, we do not need to mark/flag it. Flagging ensures that an operation does not have to travel a long \textit{chain} of backlinks. Once a link is flagged or marked it can not be a \textit{point of injection} of a \textit{new} {\sc Add} or {\sc Remove}. However, in the conflict between a {\sc Remove} of a category 3 node, and therefore shifting its predecessor, and a concurrent {\sc Remove} of the predecessor node itself, we give priority to the former. Also to guarantee a  single pointer travelling for recovery from failure due to a concurrent modify operation, we use a \textit{prelink} at each node. It connects a node to its order-node before any of the outgoing pointers are marked. The flagging and marking are performed in a definite order to avoid a malformed structure of the BST and to ensure required priorities between conflicting operations. The flag-mark order of links for a category 3 node is as following - (I) flag the incoming order-link, (II) set the prelink, (III) mark the outgoing right-link, (IV) flag the parent-link of the predecessor incoming to that, (V) flag the incoming parent-link, (VI) mark the outgoing left-link and finally (VII) mark the outgoing left-link of the predecessor. For a node belonging to category 1 or 2, because there is no node between its order-node and itself, steps (IV), (VI) and (VII) do not happen. Having performed the flagging and marking we update the flagged links according to the category of the node as described before. See Fig. \ref{bst_del}.

\indent Because we follow orderly the modifications of the links, it never allows a node to be missed by a traversal in the BST unless both its incoming links are pointed away. However, because a node may shift ``upward'' to replace its successor, the interval associated with its successor, may shift ``rightward'' i.e. to the right subtree of the node after the successor is removed. Therefore, in order to terminate a traversal at the correct location, we have the stopping criterion given in Condition \ref{cond1}. It follows from the fact that the threaded left-link of a node is connected to the node itself and a traversal in the BST uses the order in the equivalent list.
\begin{cond}\label{cond1}
Let $k$ be the search key and $k_{curr}$ be the key of the current node in the search path. If $(k~=~k_{curr})$ then stop. Else, if the next link is a threaded left-link then stop. Else, if the next link is a threaded right-link and the next node has key $k_{next}$ then check if  $k<k_{next}$. If true then stop, else continue.
\end{cond}

\indent This stopping criterion not only solves the problem of synchronization between a concurrent {\sc Remove} and a traversal for a predecessor query but also enables to achieve bound on the length of the traversal path. We shall explain that in section \ref{secAnalysis}. A similar stopping criterion is used in~\cite{Drachsler_2014_PCB} for conventional doubly-threaded BSTs.

\indent We can observe that in our BST design, two modify operations that need to change two disjoint memory-words have significantly improved conditions for progress. For example, to {\sc Remove} a category 2 node the left-link of its predecessor is never marked or flagged, therefore when such a node goes under {\sc Remove}, a concurrent injection of {\sc Add} or {\sc Remove} at the predecessor is possible. Also, because of the orderly flagging and marking of the links in more than one atomic steps, even after {\sc Remove} of the node has commenced, a link at it that comes late in the order of flagging-marking, can be modified for injection of an {\sc Add} or a {\sc Remove} of another node. These progress conditions are not possible in the existing algorithms that use ``node holding''~\cite{Ellen:2010:NBS:1835698.1835736, Howley:2012:NIB:2312005.2312036}. It shows that our algorithm has improved disjoint-access-parallelism.
\subsection{The Implementation}
\begin{algorithm}[h]
\scriptsize{
\tdfstr \nd \{\\\label{decBegin}
\Indp \el $k$\;
\tcp*[h]{Three booleans overlap with three unused bits of the pointer $ref$. left=child[0], right=child[1].}\\
\{\nptr $ref$, \bl $flag$, \bl $mark$, \bl $thread$\} child[2]\;\label{nodepointer}
\nptr $backLink$\;
\nptr $preLink$\;
\Indm\} *\nptr\; \label{decEnd}
\BlankLine
\tcp*[h]{Global variable with the fixed value of members.}\\
\nd $root[2]$ = $\{\{-\infty,~\{(\&root[0],~0,~0,~1),~(\&root[1],~0,~0,~1)\},~\&root[1],~\nul\}$, $~\{\infty,~\{(\&root[0],~0,~0,~0),~(\nul,~0,~0,~1)\},~\nul,~\nul\}\}$\; \label{gv}
}
\end{algorithm}
\indent We consider our concurrent setup as a shared memory machine in which processes are fully asynchronous and arbitrary failure halting is allowed. The read and write of a single memory-word is guaranteed to be atomic. The system provides atomic single-word \texttt{CAS} primitives. \texttt{CAS}$(R,~old,~new)$ returns \tru, if $(old=R)$, after updating $R$ to $new$, else it returns \fal. We steal three bits from a pointer, for (a) thread indicator, (b) mark-bit and (c) flag-bit, see line \ref{nodepointer}. This is easily possible in many high level languages, including C/C\verb!++!. We assume our algorithm be implemented with a safe lock-free memory reclamation scheme like hazard-pointer~\cite{michael2004hazard}.

\indent A typical node $x$ in our BST implementation is represented by a structure consisting of five memory-words corresponding to - (a) a key $k$, (b) array $child[2]$ containing two child pointers : $child[0]~:=~left(x)$ and $child[1]~:=~right(x)$, (c) a $backLink$ and (d) a $preLink$.
The bit sequence corresponding to boolean variables overlapping three stolen bits from a link is represented as $(f,~m,~t)$. One or more bits together can be set or unset using a single atomic \texttt{CAS} operation over the pointer containing it. The node structure is shown in lines \ref{decBegin} to \ref{decEnd}. We use two dummy nodes as global variables represented by a two member array of node-pointers $root$. The keys $-\infty$ and $\infty$ are stored in the two members of $root$ and they can never be deleted. Node $root[0]$ is left-child and predecessor of the node $root[1]$, see line \ref{gv} and figure \ref{bst_node_category} (c).

\subsubsection{Locating a Node}
\begin{algorithm}[h]
\scriptsize{
\tcp*[h]{{\sc Locate} returns 2 if key is found otherwise 0 or 1 if the last visited link is left or right respectively.}\\
\integer {\sc Locate}(\nptr{\hspace{-0.3em}\&}~\pre,~\nptr{\hspace{-0.3em}\&} \cur,~\el \e)\\
{\label{locateBeg}
\While{true}
{
    \tcp*[h]{Function cmp returns 2 := equal, 1 := greater than and 0 := less than}\\
    $dir$ = \cmp\hspace{-0.3em}(\e,~\cur${\spsym}e$)\;\label{cmp}

    \lIf{$(dir~{=}~2)$} 
    {\label{locsuc}
        \ret $dir$\;
    }
    \Else
    {\label{stopCriterian}
        $(R,~*,~m,~t)$ = $curr{\psym}child[dir]$\;
        \BlankLine
        \tcc*[h]{If eager helping is required then help cleaning the current node if its right child pointer has been marked; this cleaning can not fail.}\\
        \If{$((m~=~1)$ \an $(dir~=~1))$}
        {\label{eagerhelp1}
            $newPrev$ = \pre${\spsym}backLink$\;
            {\sc CleanMarked}$(prev,~curr,~dir)$\;
            \pre~=~$newPrev$\;
            $pDir$~=\cmp~\hspace{-0.3em}(\e,~\pre${\spsym}\e)$\;
            \cur~=~$($\pre${\spsym}child[pDir]){\cdot}ref$\;
            \cont\;\label{eagerhelp2}
        }

        \If{$t$}
        {\label{term2}
            \tcp*[h]{Check the stopping criterion.}\\
            $nextE~=~R{\psym}$\e\;
            \uIf{$((dir~=~0)$ \orr $($\e $< nextE))$}
            {\ret $dir$\;}
            \lElse
            {
            \pre = \cur; \cur = $R$\;
            }
        }
    }
    }
}
}
\end{algorithm}

The set operations - {\sc Contains}, {\sc Add} and {\sc Remove}, need to perform a predecessor query using a given key $k$ to locate an interval $[k_{i},~k_{j}]$, s.t. either $\{k_{i}{\leq}k{<} k_{j}\}$ or $\{k_{i}{\geq} k {>} k_{j}\}$, where $x(k_{i})$ and $x(k_{j})$ are two nodes in the BST. The function {\sc Locate} is used for that, which starts from a specified set of two consecutive nodes \{\pre, \cur\hspace{-0.3em}\} and follows the symmetric order of the internal BST to terminate at such a location $\{x(k_{prev}),~x(k_{curr})\}$. The return value of {\sc Locate} can be 0, 1 or 2 depending on whether the key $k$ is less than, greater than or equal to the key $k_{curr}$ at the termination point, line \ref{cmp}. If $k_{curr}~\neq~k$ then the desired interval is associated with the threaded outgoing link from $x(k_{curr})$ in the direction indicated by return value of {\sc Locate}, 0 denotes left and 1 denotes right. The termination criterion for {\sc Locate} implements Condition \ref{cond1}.
\begin{algorithm}[h]

\scriptsize{
\bl {\sc Contains}(\el \e)\\
{\label{searchBeg}
\tcp*[h]{Initialize the location variables with the global variables.}\\
$prev$ = $\&root[1]$; $curr$ = $\&root[0]$\;
$dir$ = {\sc Locate}~\hspace{-0.3em}$(prev,~curr,$~\e\hspace{-0.3em}$)$\;
\lIf{$(dir~=~2)$}
 {\label{chkLogDel}
       \ret \tru\;
 }
\lElse
{
       \ret \fal\;
}\label{searchEnd}
}
}

\end{algorithm}

\indent As we mentioned before, we can make a traversal eagerly help pending {\sc Remove} operations, even though it is not obstructed by them, in the situations in which proportion of {\sc Remove} increases. If that is done then a traversal cleans a node whose marked right-link it encounters during execution by calling the function {\sc CleanMark}. This functionality can be enabled using a boolean variable as an input argument to every Set operation which is further passed to the {\sc Locate} that it performs. This is not done if the implementation does not demand such eager helping. See line \ref{eagerhelp1} to \ref{eagerhelp2}. To perform a {\sc Contains} operation of key $k$, we start from the location $\{x(\infty),~x(-\infty)\}$ represented by the global variables. Having performed {\sc Locate}, the return value itself indicates whether the key $k$ was present in the BST at the point of termination of {\sc Locate}. If {\sc Locate} returns 2 then {\sc Contains} returns \tru otherwise \fal, line \ref{searchBeg} to \ref{searchEnd}.

\subsubsection{Remove operation}
\begin{algorithm}[h]

\scriptsize{
    \bl {\sc Remove}(\el \e)\\
    {\label{remBeg}
        \tcp*[h]{Initialize the location variables as before.}\\
        $prev$ = $\&root[1]$; $curr$ = $\&root[0]$\;
        $dir$ = {\sc Locate}~\hspace{-0.3em}$(prev,~curr,$~\e\hspace{-0.3em}$-~\epsilon)$;\label{locDel}\tcp*[f]{locate order-link}\\
        $(next,~f,~*,~t)$ = $curr{\psym}child[dir]$\;

        \lIf{$($\e$~\neq~next{\psym}k)$}
         {
               \ret \fal\;
         }
        \Else
        {
            \tcp*[h]{flag the order-link}\\
            $result$ = {\sc TryFlag}$(curr,~next,~prev,~$\tru\hspace{-0.3em}$)$\;\label{algtryflag}
             \If{$(prev{\psym}child[dir].ref = curr)$}
             {
                {\sc CleanFlag}$(curr,~next,~prev,~$\tru\hspace{-0.3em}$)$\;\label{algcleanflag_from_rem}
             }

        }
    \ret $result$\;
}
}
\end{algorithm}

To {\sc Remove} a node corresponding to key $k$, starting from $\{x(\infty),~x(-\infty)\}$ we locate the link corresponding to the interval containing the key $(k-\epsilon)$ line \ref{locDel}. If the {\sc Locate} terminates at $\{x(k_{prev}),~x(k_{curr})\}$ then $suc(x(k_{curr}))$ is the desired node to remove if $k$ matches with its key. $suc(x(k_{curr})$ is the node pointed by the threaded right-link of $x(k_{curr})$ and this link is indeed the order-link of $x(k)$. If $x(k)$ is located then we try to flag its order-link using {\sc TryFlag}, line \ref{algtryflag}, in order to perform the step (I) of {\sc Remove}.
\begin{algorithm}[h]
\scriptsize
{
    \bl {\sc TryFlag}(\nptr{\hspace{-0.3em}\&}~\pre,~\nptr{\hspace{-0.3em}\&} \cur,~\nptr{\hspace{-0.3em}\&} \back, \bl \isth)\\\label{tryflagbegin}
    {
        \While{true}
        {
            $pDir$~=~\cmp~\hspace{-0.3em}$($\cur${\spsym}$\e,~\pre${\spsym}$\e\hspace{-0.3em}$)$~$\&~1$\tcp*[r]{If \cmp returns 2 then \cur is the left link of \pre; so $pDir$ is changed to 0.}

            \tcp*[h]{Try atomically flagging the parent link.}\\
            $t~=~$\isth\;
            $result$ = \texttt{CAS}$($\pre${\spsym}child[$\pdir$\hspace{-0.3em}],~($\cur\hspace{-0.3em}$,0,0,t),~($\cur\hspace{-0.3em}$,1,0,t))$\;\label{atomicflag}
            \lIf{$result$}
            {
                \ret \tru\;
            }
            \Else
            {
                \tcc*[h]{The \texttt{CAS} fails, check if the link has been marked, flagged or the curr node got deleted. If flagged, return false; if marked, first clean it; else just proceed}\\
                $(newR,~f,~m,~t)$ = \pre${\spsym}child[$\pdir$\hspace{-0.3em}]$\;
                \If{$(newR~=~$\cur$\hspace{-0.3em})$}
                {
                    \lIf{$f$}
                    {
                        \ret \fal\;\label{tryflagretfal}
                    }
                    \ElseIf{$m$}
                    {\label{flaghepsclean}
                          {\sc CleanMarked}$($\pre,$~pDir)$;
                    }
                    \pre~=~\back\;
                    $pDir$~=\cmp~\hspace{-0.3em}$($\cur${\spsym}$\e,~\pre${\spsym}$\e\hspace{-0.3em}$)$\;
                    $newCurr$~=~$($\pre${\spsym}child[newPDir]){\cdot}ref$;
                    {\sc Locate}~\hspace{-0.3em}$($\pre~\hspace{-0.3em}$,~newCurr,$~\cur${\spsym}$\e\hspace{-0.3em}$)$\;
                    \If{$(newCurr\neq$\cur$)$}
                    {
                        \ret \fal\;\label{tryflagret}
                    }
                    \back = \pre${\spsym}backLink$\;\label{tryflagend}

                }
            }
        }
    }
}
\end{algorithm}

\indent {\sc TryFlag}, line \ref{tryflagbegin} to \ref{tryflagend}, returns \tru only if the operation performing it could successfully flag the desired link at line \ref{atomicflag}. If the \texttt{CAS} step to atomically flag the link fails then it checks the reason of failure. If it fails because some other thread already had successfully flagged the link, \fal is returned, line \ref{tryflagretfal}. If it fails because the link was marked then it first helps the operation that marked the link, line \ref{flaghepsclean}. It could also fail to flag a threaded link because of an {\sc Add} of a new node. In both these cases it moves a step back and locates the node whose parent-link was desired to be flagged. The address for starting the recovery is saved in the variable \back. If the node is not located then it implies that some other thread already removed the target node and so {\sc TryFlag} just returns \fal, line \ref{tryflagret}.

\begin{algorithm}[h]
\scriptsize{
    \vd {\sc TryMark}(\nptr{\hspace{-0.3em}\&}~\cur,~\integer \dir)\\
    {\label{tmbeg}

        \While{true}
        {
            $back$ = \cur${\spsym}backLink$\;
            $(next,~f,~m,~t)$ = \cur${\spsym}child[$\dir$\hspace{-0.3em}]$\;
            \lIf{$(m~==~1)$}
            {
                \brk\;
            }
            \ElseIf{$(f~==~1)$}
            {
                \If{$(t~==~0)$~~~}
                {
                    {\sc CleanFlag}$($\cur,$~next,~back,~$\fal\hspace{-0.3em}$)$;
                    \cont\;
                }
                \ElseIf{$((t~==~1)~$\an$($\dir$~==~1))$}
                {\label{trymark_ret_left}
                    {\sc CleanFlag}$($\cur,$~next,~back,~$\tru\hspace{-0.3em}$)$;
                    \cont\;
                }
            }
            $result$ = \texttt{CAS}$($\cur${\spsym}child[$\dir$\hspace{-0.3em}],$ $(next,~0,~0,~t), (next,$ $0,~1,~t))$\tcp*[r]{Try atomically marking the child link.}
            \lIf{$result$}
            {
                \brk\;\label{tmend}
            }
         }
    }
}
\end{algorithm}

\indent Having performed the {\sc TryFlag}, {\sc Remove} checks whether the target node is still there. If the target node is found then {\sc Remove} goes to clean the flag at the order-link of the target node, using the function {\sc CleanFlag}, line \ref{algcleanflag_from_rem}. The function {\sc CleanFlag} is also used to clean the flag of the parent-link of a node, and therefore a boolean variable is passed to it to inform about the thread-bit of the flagged link. If the link is an order-link then, for all categories of nodes, the next step is to set the prelink and then mark the right-link, see lines \ref{mark_cf_start} to \ref{mark_cf_end}. If the \texttt{CAS} to mark the right-link fails because of its flagging, and if the right-link is threaded then it indicates that the node itself is being shifted to replace its successor by a concurrent {\sc Remove} operation. We give priority to the shifting operation and therefore before proceeding it helps. As before, we need to recover from failures and so a node \back stores the address to restart the recovery from. Here one has to be careful that if \back was pointed at the successor node under remove then we need to change it before going to help, see line \ref{shift_back}. Having marked the right-link, {\sc Remove} proceeds to take further steps in the function {\sc CleanMark}. Also, before marking the right-link, the prelink gets pointed to the correct order-node, line \ref{setprelink}.

\begin{algorithm}[h]
\scriptsize
{
    \vd {\sc CleanFlag}(\nptr{\hspace{-0.3em}\&}~\pre,~\nptr{\hspace{-0.3em}\&} \cur,~\nptr{\hspace{-0.3em}\&} \back, \bl \isth)\\
    {
        \uIf{$($\isth$)$}
        {\label{mark_cf_start}\tcp*[h]{Cleaning a flagged order-link}\\
            \While{\tru}
            {\label{while_loop_for_right_mark}
            \tcp*[h]{To mark the right-child link}\\
                  $(next,~f,~m,~t)$ = \cur${\spsym}child[1]$\;
                  \lIf{$(m)$}
                  {
                    \brk\;
                  }
                  \ElseIf{$(f)$}
                  {
            			\tcp*[h]{Help cleaning if right-child is flagged}\\
                          \If{$($\back$=~next)$}
                          {\label{shift_back}\tcp*[h]{If \back is getting deleted then move back.}\\
                            \back$=~$\back${\spsym}backLink$\;
                          }
                          $backNode~=~$\cur${\spsym}backLink$\;
                           {\sc CleanFlag}$($\cur,$~next,~backNode,t$$)$\;
                          \If{$($\back$=~next)$}
                          {
                               $pDir$~=\cmp~\hspace{-0.3em}$($\pre${\spsym}$\e,~$backNode{\psym}$\e$\hspace{-0.3em})$\;
                                \pre~$=~$\back${\spsym}child[pDir]$\;
                          }
                  }
                  \Else
                  {
                            \lIf{$($\cur${\spsym}preLink~\neq$ \pre\hspace{-0.3em}$)$} {\cur${\spsym}preLink$ = \pre\;\label{setprelink} }
                            $result$ = \texttt{CAS}$($\cur${\spsym}child[1]$, $(next,~0,~0,~t)$, $(next,~0,~1,~t))$\label{atomicmark} \tcp*[r]{Try marking the child link.}
                            \lIf{$result$}
                            {\label{mark_cf_end}
                                \brk\;
                            }
                  }

            }
            {\sc CleanMark}$($\cur,~1$)$\;
        }
        \Else(\tcp*[f]{Cleaning a flagged parent-link})
        {
            $(right,~rF,~rM,~rT)$ = \cur${\spsym}child[1]$\;
            \uIf(\tcp*[f]{The node is to be deleted itself}){$(rM)$}
            {
                $(left,~lF,~lM,~lT)$ = \cur${\spsym}child[0]$\;
                $preNode$ = \cur${\spsym}preLink$\;
                \uIf(\tcp*[h]{A category 3 node}){$(left~\neq~preNode)$}
                {\label{tm_left}
                    {\sc TryMark}$($\cur\hspace{-0.3em}$,~0)$\;
                    {\sc CleanMark}$($\cur\hspace{-0.3em}$,~0)$\;
                }
                \Else(\tcp*[f]{This is a category 1 or 2 node})
                {\label{cf_check_category}
                    $pDir$~=\cmp~\hspace{-0.3em}$($\cur${\spsym}$\e,\pre${\spsym}$\e$\hspace{-0.3em})$\;
                    \uIf(\tcp*[f]{A category 1 node}){$(left~=~$\cur\hspace{-0.3em}$)$}
                    {
                        \texttt{CAS}$($\pre${\spsym}child[pDir], ($\cur$,~f,~0,~0), (right,~0,~0,~rT))$\;\label{ptr_swapping_cat1_1}
                        \lIf{$(!rT)$}{\label{ptr_swapping_cat1_2}\texttt{CAS}$(right{\psym}backLink,~($\cur\hspace{-0.3em}$,~0,~0,~0),~($\pre\hspace{-0.3em}$,~0,~0,~0))$\;}
                    }
                    \Else(\tcp*[f]{A category 2 node})
                    {
                        \texttt{CAS}$(preNode{\psym}child[1], ($\cur\hspace{-0.3em}$,~1,~0,~1), (right,~0,~0,~rT))$\;\label{ptr_swapping_cat2_1}
                        \lIf{$(!rT)$}{\texttt{CAS}$(right{\psym}backLink,~($\cur\hspace{-0.3em}$,~0,~0,~0),~($\pre\hspace{-0.3em}$,~0,~0,~0))$\;\label{bk_update1}}
                        \texttt{CAS}$($\pre${\spsym}child[pDir],~($\cur\hspace{-0.3em}$,~1,~0,~0),~(preNode,~0,~0,~rT))$\;
                        \texttt{CAS}$(preNode{\psym}backLink,~($\cur\hspace{-0.3em}$,~0,~0,~0),~($\pre\hspace{-0.3em}$,~0,~0,~0))$\;\label{ptr_swapping_cat2_2}
                    }
                }
            }
            \ElseIf{$(rt~$\an$rF)$}
            {\label{step6_1}\tcp*[h]{The node is moving to replace its successor}\\
                  $delNode~=~right$\;
                 \While{\tru}
                  {
                        $parent~=~delNode{\psym}backLink$\;
                        $pDir$~=\cmp~\hspace{-0.3em}$($\cur${\spsym}$\e,~\pre${\spsym}$\e$\hspace{-0.3em})$\;
                        $(*,~pF,~pM,~pT)$ = $parent{\psym}child[pDir]$\;
                        \lIf{$(pM)$}
                          {
                            {\sc CleanMark}$(parent,~pDir)$\;
                          }
                        \lElseIf{$(pF)$}
                        {
                            \brk\;
                        }
                        \ElseIf{$($\texttt{CAS}$(parent{\psym}child[pDir],~($\cur\hspace{-0.3em}$,~0,~0,~0),~($\cur\hspace{-0.3em}$,~1,~0,~0)))$}
                        {
                            \brk\;
                        }
                  }
                 $backNode~=~parent{\psym}backLink$\;
                 {\sc CleanFlag}$(parent,~$\cur\hspace{-0.3em}$,~backNode,~$\tru\hspace{-0.3em}$)$\;\label{step6_2}
            }
        }
    }
}
\end{algorithm}

\indent If {\sc CleanFlag} is performed on a link that is not threaded then it is helping an operation that has flagged the parent of a node or of its predecessor. In that case it is determined whether the link is the parent-link of a node under {\sc Remove} or that of its predecessor by checking the flag, mark and thread bits of the right-link. The right-link of a node being removed is marked, whereas that for the predecessor of a node being removed is always flagged and threaded. Accordingly, either step (V) (lines \ref{step6_1} to \ref{step6_2}) or step (VI) (lines \ref{tm_left}) are performed. To perform the step (VI) the outgoing left-link of a category 3 node is marked, and for that the function {\sc TryMark} is used. {\sc TryMark} atomically inserts the mark-bit at a link specified by its direction outgoing from a node, see lines \ref{tmbeg} to \ref{tmend}. Note that, if the \texttt{CAS} to atomically mark a link fails due to a concurrent flagging of the link then {\sc TryMark} helps the concurrent operation that flagged the link, except in the case when the link is a threaded left-link. This is because a flagged and threaded left-link of a node indicates that the node is a category 1 node under remove and we give priority to the operation that must have already flagged its right link to shift it.

\indent Because the last step for category 1 and category 2 nodes is flagging their parent-link before the pointers are appropriately swapped, {\sc CleanFlag} also includes the final swapping of pointers for such nodes. Having determined that a node belongs to category 1 or 2 by checking the equality of left-node and order-node at line \ref{cf_check_category}, it performs the pointer swapping for category 1 nodes in lines \ref{ptr_swapping_cat1_1} and \ref{ptr_swapping_cat1_2}. Pointer swapping steps to clean a category 2 node is shown between lines \ref{ptr_swapping_cat2_1} and \ref{ptr_swapping_cat2_2}.

\begin{algorithm}[h]
\scriptsize{
    \vd {\sc CleanMark}(\nptr{\hspace{-0.3em}\&}~\cur,~\integer \mdir)\\
    {\label{cmbegin}
    $(left,~lF,~lM,~lT)$ = \cur${\spsym}child[0]$\;
    $(right,~rF,~rM,~rT)$ = \cur${\spsym}child[1]$\;

    \uIf{$($\mdir$~==~1)$}
    {\tcc*[h]{The node is getting deleted itself.  if it is category 1 or 2 node, flag the incoming parent link;  if it is a category 3 node, flag the incoming parent link of its predecessor.}\\

        \While{\tru}
         {\label{conbackl}
                $preNode~=~delNode{\psym}preLink$\;
                \uIf(\tcp*[f]{Category 1 or 2 node.}){$(preNode~==~left)$}
                  {
                        $parent~=~delNode{\psym}backLink$\;
                        $back~=~parent{\psym}backLink$\;
                        {\sc TryFlag}$(parent,~$\cur\hspace{-0.3em}$,~back,~$\tru\hspace{-0.3em}$)$\;
                        \If{$(parent{\psym}child[pDir].ref~==~$\cur\hspace{-0.3em}$)$}
                        {
                            {\sc CleanFlag}$(parent,~$\cur\hspace{-0.3em}$,~back,~$\tru\hspace{-0.3em}$)$\;
                            \brk\;
                        }
                  }
                \Else
                {\tcp*[h]{Category 3 node.}\\

                        $preParent~=~preNode{\psym}backLink$\;
                        $(*,~pPF,~pPM,~pPT)$ = $preParent{\psym}child[1]$\;
                        $backNode~=~preParent{\psym}backLink$\;
                        \lIf{$(pM)$}
                          {
                            {\sc CleanMark}$(preParent,~1)$\;
                          }
                        \ElseIf{$(pF)$}
                        {
                            {\sc CleanFlag}$(preParent,~preNode,~backNode,~$\tru\hspace{-0.3em}$)$;\brk\;
                        }
                        \ElseIf{$($\texttt{CAS}$(parent{\psym}child[pDir], ($\cur\hspace{-0.3em}$,~0,~0,~0), ($\cur\hspace{-0.3em}$,~1,~0,~0)))$}
                        {
                            {\sc CleanFlag}$(preParent,~preNode,~backNode,~$\tru\hspace{-0.3em}$)$;\brk\;\label{conback2}
                        }

                }
          }

    }
    \Else
    {\tcp*[h]{The node is getting deleted ($\dagger$) or moved to replace its successor ($\ddagger$).}\\

        \uIf{$(rM)$}
        {\tcp*[h]{($\dagger$) clean its left marked link.}\\\label{mark_leftPlink}
            $preNode~=~$\cur${\spsym}preLink$\;
            {\sc TryMark}$(preNode,~0)$\;
            {\sc CleanMark}$(preNode,~0)$\;
        }
        \ElseIf{$(rt~$\an$rF)$}
        {
            \tcp*[h]{($\ddagger$) change links accordingly}\\
            $delNode$ = $right$\;\label{ptr_swapping_cat3_1}
            $delNodePa$ = $delNode{\psym}backLink$\;
            $preParent$ = \cur${\spsym}backLink$\;
            $pDir$~=\cmp~\hspace{-0.3em}$(delNode{\spsym}$\e,~$delNodePa{\psym}$\e$\hspace{-0.3em})$\;
            $(delNodeL,*,*,~dlT)$ = $delNode{\psym}child[0]$\;
            $(delNodeR,*,*,~drT)$ = $delNode{\psym}child[1]$\;
            \texttt{CAS}$(preParent{\psym}child[1],~($\cur$\hspace{-0.3em},f,0,0)$,$~(left,lF,0,lT))$\;\label{flag_copy}

            \lIf{$(!lT)$}{\texttt{CAS}$(left{\psym}backLink,~($\cur$\hspace{-0.3em},0,0,0),$ $(preParent,0,0,0))$\;\label{bk_update2}}

            \texttt{CAS}$($\cur${\spsym}child[0],~(left,0,1,lT),(delNodeL,0,0,0))$\;
            {\texttt{CAS}$(delNodeL{\psym}backLink,~(delNode,0,0,0),~($\cur$\hspace{-0.3em},0,0,0))$\;\label{bk_update3}}
            \texttt{CAS}$($\cur${\spsym}child[1],(right,1,0,1),(delNodeR,0,0,drT))$\;
            \lIf{$(!drT)$}{\texttt{CAS}$(delNodeR{\psym}backLink,$ $(delNode,0,0,0),$ $($\cur$\hspace{-0.3em},0,0,0))$\;\label{bk_update4}}
            \texttt{CAS}$(delNodePa{\psym}child[pDir],~(delNode,1,0,0),~($\cur$\hspace{-0.3em},0,0,0))$\;\label{li_remove}
            {\texttt{CAS}$($\cur${\spsym}backLink,~(preParent,0,0,0),$ $(delNodePa,0,0,0))$\;\label{cmend}}
        }

    }
}
}
\end{algorithm}

\indent The {\sc CleanMark} function is used to help an operation that puts a mark-bit at a link. Depending on whether the link is a left- or a right-link, it takes different steps. If the link happens to be a right-link then, for category 1 and 2 nodes, the final step of flagging the parent is executed; otherwise for a category 3 node the parent of its predecessor is flagged. To determine the category of a node, its order-node is checked if it coincides with its left-child. If it is a category 3 node then {\sc CleanMark} attempts to flag the parent of the order-node. On failing because a concurrent operation had marked the link, it helps that operation. However, after helping it may be that the node under {\sc Remove} changes from a category 3 to a category 2 node, so its status is checked again, see lines \ref{conbackl} to \ref{conback2}.

\indent A mark-bit at a left-link indicates that it is either coming out from a category 3 node or its predecessor. In the first case, the final step i.e. step (VII) to mark the left-link of predecessor of a category 3 node is performed, line \ref{mark_leftPlink}. In the second case the pointer swapping steps of a category 3 nodes are performed, see lines \ref{ptr_swapping_cat3_1} to \ref{cmend}.

\indent Return value of {\sc Remove} is that of the absolutely first {\sc TryFlag} that it performs if the node was located.

\subsubsection{Add operation}
\indent To {\sc Add} a new data-node with key $k$ in the BST, we {\sc Locate} the target interval $[k_{i},~k_{j}]$, associated with a threaded link, containing key $k$. If {\sc Locate} returns 2 then key is present in the BST and therefore {\sc Add} returns \fal. If it returns 0 or 1 then we create a new node containing the key $k$. Its left-link is threaded and connected to itself and right-link takes the value of the link to which it needs to be connected, line \ref{lem5_2}. Note that when a new node is added, both its children links are threaded. Also, its backlink is pointed to the node $x(k_{i})$. The link which it needs to connect to is modified in one atomic step to point to the new node using a \texttt{CAS}. If the \texttt{CAS} succeeds then \tru is returned. On failure, it is checked whether the target link was flagged, marked or another {\sc Add} operation succeeded to insert a new node after we read the link. In case a new node is inserted, we start locating for a new proper link starting with the location comprising nodes at the two ends of the changed link. On failure due to marking or flagging of the current link, the concurrent {\sc Remove} operation is helped. Recovery from failure due to a flagging or marking by a concurrent {\sc Remove} operation makes it to go one link back following the backlink of \pre. After locating the new proper link, {\sc Add} is retried.
\begin{algorithm}[h]

\scriptsize{
    \bl {\sc Add}(\el \e)\\
    {
        $prev$ = $\&root[1]$; $curr$ = $\&root[0]$\;
        \tcc*[h]{Initializing a new node with supplied key and
    left-link threaded and pointing to itself.}\\
        $node$ = \new \nod\hspace{-0.3em}(\e)\;
        $node{\psym}child[0]$ = $(node,~0,~0,~1)$\;\label{lem5_1}
        \BlankLine
        \While{true}
        {
            $dir$ = {\sc Locate}~\hspace{-0.3em}$(prev,~curr,$~\e\hspace{-0.3em}$)$\;
            \uIf(\tcp*[h]{key exists in the BST}){$(dir~=~2)$}
             {\label{chkAddDel}
                \ret \fal\;
             }
            \Else
            {
                $(R,~*,~*,~*)$ = $curr{\psym}child[dir]$\;
                \tcc*[h]{The located link is threaded. Set the right-link of the adding node to copy this value}\\
                $node{\psym}child[1]$ = $(R,~0,~0,~1)$\;\label{lem5_2}
                $node{\psym}backLink$ = $curr$\;\label{lem8_1}
                $result$ = \texttt{CAS}$(curr{\psym}child[dir],~(R,~0,~0,~1),$ $~(node,~0,~0,~0))$\label{addsuc}\tcp*[r]{Try inserting the new node.}
                \BlankLine
                \lIf{$result$}
                {
                    \ret \tru\;
                }
                \Else
                {
                    \tcc*[h]{If the \texttt{CAS} fails, check if the link has been marked, flagged or a new node has been inserted. If marked or flagged, first help cleaning.}\\
                    $(newR,~f,~m,~t)$ = $curr{\psym}child[dir]$\;
                    \BlankLine
                    \If{$(newR~=~R)$}
                    {
                        $newCurr$ = $prev$\;
                        \lIf{$m$}
                        {
                            {\sc CleanMarked}$(curr,~dir)$\;
                        }
                        \ElseIf{$f$}
                        {
                            {\sc CleanFlagged}$(curr,~R,~prev,~$\tru\hspace{-0.3em}$)$\;
                        }
                        $curr$ = $newCurr$\;
                        $prev$ = $newCurr{\psym}backLink$\;
                    }
                }
                \BlankLine
            }
            \BlankLine
        }
    }
}

\end{algorithm}
\section{Correctness and Complexity}\label{secAnalysis}
In this section we first show that executions in our algorithm produce a correct BST with linearizable operations. Then we prove the lock-free progress property and finally we discuss its amortized step complexity.

\subsection{Correctness}
Here we present a proof-sketch of correctness, due to space constraints. A detailed proof can be found in the technical report~\cite{Bapi2014BST:TR}.
First we give classification of nodes in a BST $\Upsilon$ implemented by our algorithm.

\begin{definition}\label{def1} A node $x{\in}\Upsilon$ is called logically removed if its right-link is marked and $\exists~y{\in}\Upsilon$ s.t. either $left(y)~=~x$ or $right(y)~=~x$.
\end{definition}
\vspace{-7pt}
\begin{definition} A node $x{\in}\Upsilon$ is called physically removed if $\nexists~y{\in}\Upsilon$ s.t. either $left(y)~=~x$ or $right(y)~=~x$ or $y{\psym}backLink~=~x$.
\end{definition}
\vspace{-7pt}
\begin{definition} A node $x{\in}\Upsilon$ is called regular if it is neither logically removed nor physically removed.
\end{definition}

\indent A node ever inserted in $\Upsilon$ has to fit in one of the above categories. At a point in the history of executions, the BST formed by our algorithm contains elements stored in regular nodes or logically removed nodes. Before we show that the return values of the set operations are consistent with this definition according to their linearizability, we have to show that the operations work as intended and the structure formed by the nodes operated with these operations maintains a valid BST. We present some invariants maintained by the operations and the nodes in lemmas \ref{lem1} to \ref{lem12}.
\begin{lemma}\label{lem1} If a {\sc Locate}$($\pre, \cur, $k)$ returns $dir$ and terminates at $[x(k_{prev}),~x(k_{curr})]$ then \begin{enumerate}[(a)]
\item Either $k_{curr}~{\leq}~k$ or $k_{curr}~{\geq}~k$.
\item If $k_{curr}~{\neq}~k$ then the link $x(k_{curr}){\psym}child[dir]$ is threaded.
\item $x(k_{curr})$ is not physically deleted.
\end{enumerate}
\end{lemma}
\vspace{-7pt}
\begin{lemma}\label{lem2} A {\sc Contains} operation returns true if and only if the key is located at a non-physically removed node.\end{lemma}
\vspace{-7pt}
\begin{lemma}\label{lem3} An {\sc Add} always happens at an unmarked and unflagged threaded link.\end{lemma}
\vspace{-7pt}
\begin{lemma}\label{lem4} A {\sc Remove} always starts by flagging the incoming order-link to a node.\end{lemma}
\vspace{-7pt}
\begin{lemma}\label{lem5} When a node is inserted, both its left- and right- links both threaded.\end{lemma}
\vspace{-7pt}
\begin{lemma}\label{lem6} Before a node is logically removed its incoming order-link is flagged and its prelink points to its correct order-node.\end{lemma}
\vspace{-7pt}
 \begin{lemma}\label{lem8} Backlink of a node always points to a non physically removed node.\end{lemma}
\vspace{-7pt}
\begin{lemma}\label{lem9} An unthreaded left-link can not be both flagged and marked.\end{lemma}
\vspace{-7pt}
\begin{lemma}\label{lem10} A right-link can not be both flagged and marked.\end{lemma}
\vspace{-7pt}
\begin{lemma}\label{lem11} A link once marked never changes again.\end{lemma}
\vspace{-7pt}
\begin{lemma}\label{lem12} If a node gets logically removed then it will be eventually physically removed.\end{lemma}

\indent Lemma \ref{lem1} follows from the lines \ref{locsuc} and \ref{term2}. {\sc Contains} returns \tru only if {\sc Locate} returns 2 and that happens only if \cur is non-physically removed at line \ref{cmp} during its execution, this proves lemma \ref{lem2}. In the case of a key match, {\sc Add} returns \fal and otherwise it tries adding the key using an atomic \texttt{CAS}. If the \texttt{CAS} fails, it always uses {\sc Locate} to find the desired link before retrying the \texttt{CAS} to add the new node. From this observation and using \ref{lem1}(b), lemma \ref{lem3} follows. By lemma \ref{lem1} if $x(k)$ is present in the tree then {\sc Locate}$($\pre, \cur, $(k-\epsilon))$ will always terminate at a location such that \cur is order-node of $x(k)$ and that establishes lemma \ref{lem4}. Lemma \ref{lem5} follows from lines \ref{lem5_1} and \ref{lem5_2}. Line \ref{flag_copy} ensures that even if the function {\sc CleanFlag} helps a pending {\sc Remove}, before it could successfully mark the right-link at line \ref{atomicmark} the flag that was put on the order link at line \ref{algtryflag} is copied to the changed order-link. Also, the line \ref{setprelink} inside the while loop ensures that prelink is always set to the order-node. That proves the correctness of lemma \ref{lem6}. When a node is added its backlink is pointed to \cur at line \ref{lem8_1}. Before a {\sc Remove} operation returns it ensures that the backlinks of the predecessor, left-child and right-child, if present for the node under {\sc Remove}, are appropriately updated at lines \ref{ptr_swapping_cat1_2}, \ref{bk_update1}, \ref{ptr_swapping_cat2_2}, \ref{bk_update2}, \ref{bk_update3}, \ref{bk_update4} and \ref{cmend}. Hence, by induction, lemma \ref{lem8} is proved. In our algorithm we always use an atomic \texttt{CAS} to put a flag or mark-bit in a pointer. Whenever a \texttt{CAS} fails we check the possible reason. The function {\sc TryMark} helps cleaning the flag in all cases except when the link has direction 0 and it is threaded, line \ref{trymark_ret_left}. These observations prove lemmas \ref{lem9} and \ref{lem10}. Lemma \ref{lem11} follows from lemmas \ref{lem9} and \ref{lem10}. Lemma \ref{lem11} proves that once the right-link of a node is marked, it can not be reversed and if the thread invoking {\sc Remove} to mark it becomes faulty then eventually another thread invoking a possible {\sc Add} or {\sc Remove} which gets obstructed, will help to complete the physical removal. Having proved the lemmas listed above, it is trivial to observe that whenever a pointer is dereferenced it is not null. And, by the initialization of the global variable to $\infty$ and $-\infty$ , at line \ref{gv}, the two starting nodes are never deleted.\\
\indent Hence, we state proposition \ref{treedef} whose proof will follow by the above stated lemmas and the fact that a thread always takes a correct ``turn'' during traversal according to the symmetric order of the internal BST.
\begin{proposition}\label{treedef}
The union of the regular and logically removed nodes operated under the set operations in the algorithm Efficient Lock Free BST maintain a valid internal Binary Search Tree.
\end{proposition}
\indent An execution history in our implementation may consist of {\sc Add}, {\sc Remove} and {\sc Contains} operations. We present the linearization point of the execution of these operations. Proving that a history consisting of concurrent executions of these operations is legal will be ordering these linearization points. The linearization points of the operations are as following:

\textbf{\sc Add} - For a successful {\sc Add} operation, execution of the \texttt{CAS} at line \ref{addsuc} will be the linearization point. For an unsuccessful one the linearization point will be at line \ref{cmp} where a key in the tree is found matched.

\textbf{\sc Remove} - For a successful {\sc Remove} operation the linearization point will be the successful \texttt{CAS} that swaps the flagged parent link. For an unsuccessful one there may be two cases - (a) if the node is not located then it is treated as an unsuccessful {\sc Contains} and its linearization point will be accordingly (b) if the node is located but its order-link got flagged by another concurrent {\sc Remove} then its linearization point is just after the linearization point of that {\sc Remove}.

\textbf{\sc Contains} - Our algorithm varies according to the read-write load situation. In case we go for eager helping by a thread performing {\sc Locate}, a successful {\sc Contains} shall always return a regular node. However, if we opt otherwise then a successful {\sc Contains} returns any non-physically removed node. In both situations a successful {\sc Contains} will be linearized at line \ref{cmp}. An unsuccessful one, if the node never existed in the BST, is linearized at the start point. And, if the node existed in the BST when the {\sc Contains} was invoked but got removed during its traversal by a concurrent {\sc Remove} then the linearization point will be just after the successful \texttt{CAS} operation that physically removed the node from the BST.

\indent Following the linearization points as described above we have proposition \ref{linearize}:
\begin{proposition}\label{linearize}
The Set operations in the algorithm Efficient Lock Free BST are linearizable.
\end{proposition}

\subsection{Lock-Freedom}
The lemmas \ref{lem6}, \ref{lem9}, \ref{lem10} and \ref{lem11} imply the following lemma.
\begin{lemma}\label{rem_obstruct} If {\sc Remove}$(x)$ and {\sc Remove}$(y)$ work concurrently on nodes $x$ and $y$ then without loss of generality
\begin{enumerate}[(a)]
\item If $x$ is the left-child of $y$ and both $x$ and $y$ are logically deleted then {\sc Remove}$(x)$ finishes before {\sc Remove}$(y)$.
\vspace{-7pt}
\item If $x$ is the right-child of $y$ and both $x$ and $y$ are logically deleted then {\sc Remove}$(y)$ finishes before {\sc Remove}$(x)$.
\vspace{-7pt}
\item If $x$ is the predecessor of $y$ and the order-links of both $x$ and $y$ have been successfully flagged then {\sc Remove}$(y)$ finishes before {\sc Remove}$(x)$.
\vspace{-7pt}
\item If $x$ is the predecessor of $y$ and $x$ has been logically deleted then {\sc Remove}$(x)$ finishes before the order-link of $y$ could be successfully flagged.
\vspace{-7pt}
\item If $x$ is the left-child of the predecessor of $y$ and the incoming parent-link of $x$ has been successfully flagged then {\sc Remove}$(x)$ finishes before {\sc Remove}$(y)$.
\vspace{-7pt}
\item If $x$ is the left-child of the predecessor of $y$ and the left-link of the predecessor of $y$ has been successfully marked then {\sc Remove}$(y)$ finishes before {\sc Remove}$(x)$.
\vspace{-7pt}
\item In all other cases {\sc Remove}$(x)$ and {\sc Remove}$(y)$ do not obstruct each other.
\end{enumerate}
\end{lemma}

By the description of our algorithm, a non-faulty thread performing {\sc Contains} will always return unless its search path keeps on getting longer forever. If that happens, an infinite number of {\sc Add} operations would have successfully completed adding new nodes making the implementation lock-free. So, it will suffice to prove that the modify operations are lock-free. Considering a thread $t$ performing a pending operation $op$ on a BST $\Upsilon$ and takes infinite steps, and, no other modify operation completes after that. Now, if no modify operation completes then the tree remains unchanged forcing $t$ to retract every time it wants to inject its own modification on $\Upsilon$. This is possible only if every time $op$ finds its injection point flagged or marked. This implies that a {\sc Remove} operation is pending. It is easy to observe in the function {\sc Add} that if it gets obstructed by a concurrent {\sc Remove} then before retrying after recovery from failure it helps the pending {\sc Remove} by taking all the remaining steps of that. Also from lemma \ref{rem_obstruct}, whenever two {\sc Remove} operations obstruct each other, one finishes before the other. It implies that whenever two modify operations obstruct each other one finishes before the other and so the tree changes. It is contrary to our assumption. Hence, by contradiction we show that no non-faulty thread shall remain taking infinite steps if no other non-faulty thread is making progress. This proves the following proposition \ref{lockfree}.
\begin{proposition}\label{lockfree}
Lock-freedom is guaranteed in the algorithm Efficient Lock Free BST.
\end{proposition}

\subsection{Complexity}
\indent Having proved that our algorithm guarantees lock-freedom, though we can not compute worst-case complexity of an operation, we can definitely derive their amortized complexity. We derive the amortized step complexity of our implementation by the accounting method along the similar lines as in~\cite{fomitchev2004lock, DBLP:conf/podc/OshmanS13}.

\indent In our algorithm, an {\sc Add} operation does not have to hold any pointer and so does not obstruct an operation. We observe that after flagging the order-link of a node, a {\sc Remove} operation $op_r$ takes only a constant number of steps to flag, mark and swap pointers connected to the node and to its predecessor, if any, in addition to setting the prelink of the node under {\sc Remove}. So, if a modify operation $op$ gets obstructed by  $op_r$ then it would have to perform only a constant number of extra steps in order to help $op_r$ provided that during the help it is not obstructed further by another {\sc Remove} operation $op^{'}_r$. This observation implies following lemma
\begin{lemma}\label{lemextrasteps} An obstructing operation $op$ makes an obstructed operation $op'$ take only a constant number of extra steps in order to finish the execution of $op'$.\end{lemma}

\indent Now for an execution $E$, let $n$ be the number of nodes at the beginning of $E$. Let $\mathcal{O}$ be the set of operations and let $\mathcal{S}$ be the set of steps taken by all $op{\in}\mathcal{O}$. Considering the invocation point $t_i(op)$ of $op$ to be the time it reads the root node, and response point $t_r(op)$ to be the time it reads or writes at the last link before it leaves the tree, its \textit{interval contention} $c_I$ is defined as the total number of operations $op'$ whose execution overlaps the interval $[t_i(op),~t_r(op)]$~\cite{afek2002long}. We define a function $f:\mathcal{S}\mapsto\mathcal{O}$ such that:
\begin{enumerate}[(a)]
\item Had there been no contention, all the essential steps $s$, representing read, write and \texttt{CAS} taken by an operation $op$ is mapped to $op$ by $f$.
\vspace{-7pt}\item In case of contention, any failed \texttt{CAS} by an operation $op$ is mapped to the operation $op'$ whose successful \texttt{CAS} causes the failure.
\vspace{-7pt}\item If an extra read is performed by a traversal due to an added node to the set of existing nodes by a concurrent {\sc Add} operation $op$ then it is mapped to $op$.
\vspace{-7pt}\item Any read, write or \texttt{CAS} step $s$ taken by an operation $op$ after the first failed \texttt{CAS} and before retrying at the same link i.e. during helping is mapped to an operation $op'$ that performed the successful \texttt{CAS} in order to make $op$ help it. This includes resetting of prelink, if needed.
\end{enumerate}

\indent Now we bound the length of the traversal path for a predecessor query in our implementation. Note that, all the set operations have to perform a predecessor query by key $k$ to {\sc Locate} an interval $[k_{i},~k_{j}]$ s.t. either $\{k_{i}\leq k<k_{j}\}$ $\{k_{i}\geq k>k_{j}\}$ and $x(k_i)$ and $x(k_j)$ are two nodes in the BST. Let us define the \textit{access-node} of an interval as the order-node of the node that the interval associates with. We define \textit{distance} of an interval from a traversal $op$ as the number of links that $op$ traverses from its current location to read the access-node of the interval. Suppose that at $t_i(op)$ there are $n$ nodes in the BST $\Upsilon$. Clearly, distance of the interval associated with any node for $op$ at  $t_i(op)$ is $O(H(n))$. It is also easy to observe that unless $op$ accesses a node $x$, distance of any node in the subtree rooted at it is $O(ht(x))$. When $op$ travels in the left subtree of a category 3 node $x$ and if it gets removed, the interval associated with it gets associated with the leftmost node in the right subtree of the node that replaces it and so the distance of that interval from $op$ changes. There is no such change for a category 1 or category 2 node. However, once an interval gets associated with the leftmost node in the right-subtree of $x$, which is obviously a category 1 node, its distance can not become more than $O(ht(x))$ from a traversal that has accessed $x$. These observations show that the path length of a traversal in our lock-free BST is bounded by $2H(n)$, if no node is added in the traversal path. In case a node is added, the extra read cost is mapped to the concurrent {\sc Add}.

\indent Having identified the operations to map for a step, it is easy to observe that an operation $op'$ to which a step $s$ by an operation $op$ is mapped, always has its execution interval overlapping $[t_i(op),~t_r(op)]$. So, because (a) the traversal path is bounded by $O(H(n))$, (b) a constant number of steps are needed for a modify operation after locating its target (node or link) and (c) by lemma \ref{lemextrasteps}, to an operation $op$ only a constant number of extra steps can be mapped by any concurrent operation $op'$ counted in $c_I$, we can conclude that the amortized step complexity of a set operation in our lock-free BST algorithm is $O(H(n) + c_I)$.

\indent As we described in the section \ref{secAlgo}, we give choice of eager helping to operations according to the read-write load. Now if that happens, we do not need to count all the operations whose executions overlap the interval $[t_i(op),~t_r(op)]$. We can then use a tighter notion of \emph{point contention} $c_P$~\cite{attiya2003algorithms}, which counts the maximum number of operations that execute concurrently at any point in $[t_i(op),~t_r(op)]$. In that case, given the above discussion, along the similar lines as presented in~\cite{fomitchev2004lock}, we can show that for any execution $E$, the average amortized step complexity of a set operation $op$ in our algorithm will be $$\hat{t}_{op{\in}E} \in O\left(\frac{\sum_{op{\in}E}(H(n(op))+c_P(op))}{|\{op\in E\}|}\right)$$ where $n(op)$ is the number of nodes in the BST at the point of invocation of $op$ and $c_P(op)$ is its point contention during $E$. That concludes the amortized analysis of our algorithm.

It is straightforward to observe that the number of memory-words used by a BST with $n$ nodes in our design is $5n$. 
\section{Conclusion and Future Work}\label{secConclude}
In this paper we proposed a novel algorithm for the implementation of a lock-free internal BST. Using amortized analysis we proved that all the operations in our implementation run in time $O(H(n) + c)$. We solved the existing problem of ``retry from scratch'' for modify operations after failure caused by a concurrent modify operation, which resulted into an amortized step complexity of $O(cH(n))$. This improvement takes care of an algorithmic design issue for which the step complexity of modify operations increases dramatically with the increase in the contention and the size of the data structure. This is an important improvement over the existing algorithms. Our algorithm also comes with improved disjoint-access-parallelism compared to similar lock-free BST algorithms. We also proposed a conservative helping technique which adapts to read-write load on the implementation. We proved the correctness showing linearizability and lock-freedom of the proposed algorithm.\\
\indent We plan to thoroughly evaluate our algorithm experimentally vis-a-vis existing concurrent set implementations. 
\section{Acknowledgement}
This work was supported by the Swedish Research Council under grant number 37252706 as part of Project SCHEME (www.scheme-project.org).
\pagebreak
\bibliographystyle{abbrv}

\end{document}